\documentclass[twocolumn,aps,prl,reprint,superscriptaddress]{revtex4-1}

\usepackage{graphicx}
\usepackage{amsmath}
\usepackage{amsfonts}
\usepackage{hyperref}
\hypersetup{pdftitle="Optofluidic control of the dispersion of nanoscale dumbbells",pdfdisplaydoctitle}

\renewcommand{\Re}{\operatorname{Re}}

\begin{document}


\title{Optofluidic control of the dispersion of nanoscale dumbbells}

\author{M.~Mel\'endez}
\email{marc.melendez@uam.es}
\affiliation{Universidad Aut\'onoma de Madrid, 28049, Madrid, Spain.}
\author{N.~Alc\'azar~Cano}
\affiliation{Universidad Aut\'onoma de Madrid, 28049, Madrid, Spain.}
\author{R.~P.~Pel\'aez}
\affiliation{Universidad Aut\'onoma de Madrid, 28049, Madrid, Spain.}
\author{J.~J.~S\'aenz}
\affiliation{Donostia International Physics Center (DIPC),
Paseo Manuel Lardizabal 4, 20018, Donostia-San Sebastian, Spain.}
\affiliation{IKERBASQUE, Basque Foundation for Science, 48013 Bilbao, Spain.}
\author{R.~Delgado-Buscalioni}
\affiliation{Universidad Aut\'onoma de Madrid, 28049, Madrid, Spain.}
\affiliation{Condensed Matter Physics Center (IFIMAC), Universidad Autónoma
de Madrid, 28049 Madrid, Spain.}

\date{today}


\begin{abstract}
Previous research has shown that gold nanoparticles immersed in water in an 
optical vortex lattice formed by the perpendicular intersection of two standing 
light waves with a $\pi/2\ \mathrm{rad}$ phase difference will experience 
enhanced dispersion that scales with the intensity of the incident laser. We 
show that flexible nanoscale dumbbells (created by attaching two such gold 
particles by means of a polymer chain) in the same field display different types 
of motion depending on the chain length and field intensity. We have not 
disregarded the secondary optical forces due to light scattering. The dumbbells 
may disperse, rotate or remain trapped. For some values of the parameters, the 
(enhanced) dispersion possesses a displacement distribution with exponential 
tails, making the motion anomalous, though Brownian.
\end{abstract}


\maketitle

Optofluidics faces the challenging problem of understanding the interactions of
light waves, electrons, and fluid and solid matter at the micro- and nanoscale.
Previous work has shown that it can provide a way to control the transport
properties of nanoscale objects, and it has already been applied to the guiding
and sorting of particles in microfluidic flows \cite{MacDonald2003,Xiao2010}.

Numerical experiments have shown how to control the magnitude of the mean square
displacement in a dilute suspension of gold nanoparticles in water by creating a
stationary  optical field at the intersection of two coherent laser beams with
wavelengths close to the plasmon resonance ($\lambda \approx 395\ \mathrm{nm}$) 
\cite{Albaladejo2009,Zapata2016}. In particular, Albaladejo \textit{et al.} 
demonstrated that perpendicular beams with a phase lag of $\pi/2\ \mathrm{rad}$ 
enhance the dispersion of nanoparticles by a factor proportional to the power 
density of the laser \cite{Albaladejo2009}.

We aim here to provide a method for tuning the dispersion properties of
nanoscale dumbbells created by attaching two identical $50\ \mathrm{nm}$-radius
gold spheres by means of a polymer strand \cite{Busson2011}. Even though the
setup in \cite{Albaladejo2009} enhances the dispersion for gold nanoparticles,
the mean square displacement for dumbbells depends critically on the field
intensity and the length of the connecting strand compared to the wavelength, as
we shall show below. We will use the term diffusion to refer to \textit{thermal}
diffusion caused by random molecular collisions, and dispersion to refer to the
combined effect of thermal fluctuations, optical forces and hydrodynamic
coupling.

\section{Simulation setup}

\begin{figure}
  \includegraphics[width = 0.8 \columnwidth]{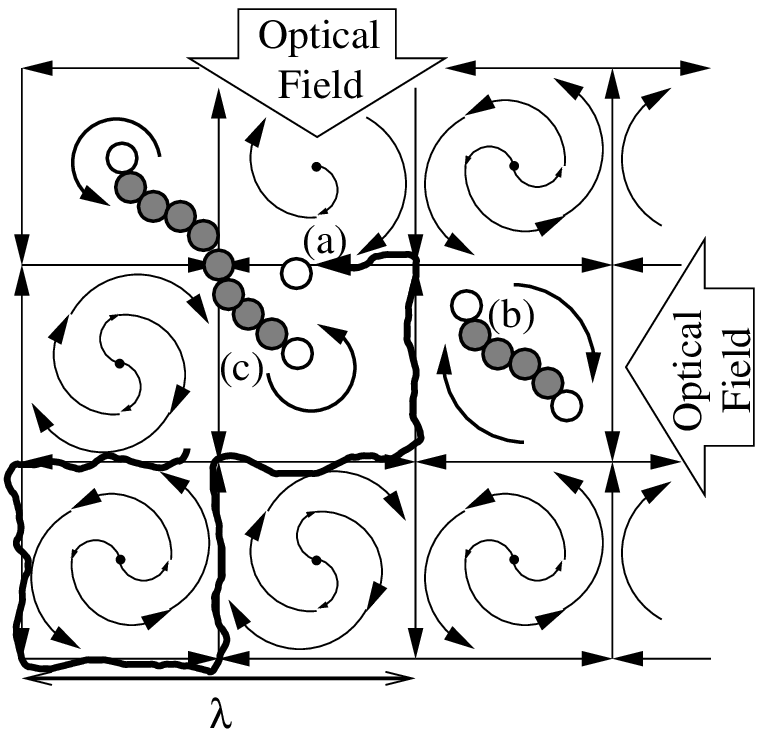}
  \includegraphics[width = 0.6 \columnwidth]{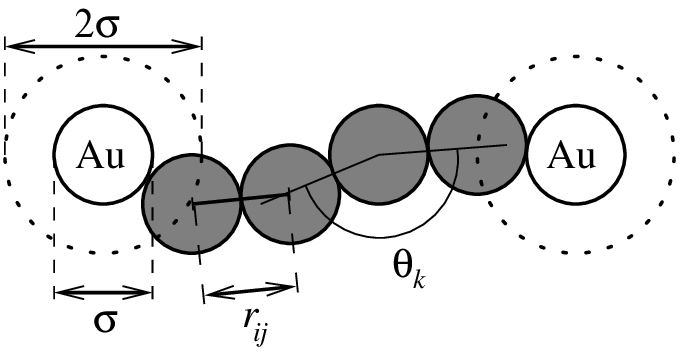}
  \caption{\label{fieldfig}Simplified representation of the forces in the
  optical vortex lattice in Eq. (\ref{force_field}). A single gold particle (a) 
  experiences enhanced dispersion. Short dumbbells follow similar trajectories. 
  We can easily trap the dumbbells (b) and (c) using one or two vortices if we 
  neglect hydrodynamic interactions. \textit{Below}: Schematic drawing of a 
  flexible dumbbell. Two gold particles attached by means of a FENE bead-spring 
  chain (which limits the interparticle separation $r_{ij} < 3 r$, where $r$ is 
  the radius of a bead) interact with an optical force field. WCA interactions 
  with diameter $\sigma$ model excluded volume effects. An extra coat of 
  transparent material (dotted lines) was included in some of the 
  simulations. Angular springs between consecutive links oppose bending and tend
  to restore the angles to $\theta_k = \pi\ \mathrm{rad}$.}
\end{figure}

Following Ref. \cite{Albaladejo2009}, we began with a nonconservative optical 
field generated in water by the intersection of two perpendicular coherent laser
beams with a $\pi/2\ \mathrm{rad}$ phase difference polarised along the $z$ 
axis. The resulting force field acting on gold nanoparticles along the $xy$
plane, $\mathbf{F}_{opt}(x,\ y)$, an optical vortex lattice (Fig.
\ref{fieldfig}), corresponded to the equation below, 
\begin{align}
\label{force_field}
  \mathbf{F}_{opt} = & 2 \alpha' \frac{n}{c} I\ \nabla(\sin(kx) + \sin(ky))^2
                         \nonumber \\
                     & + 2 \alpha'' \frac{n}{c} I\ 
                            \nabla \times (2 \cos(kx) \cos(ky)\ \mathbf{e}_z).
\end{align}

The refractive index was set to $n = \sqrt{1.8}$. We assumed that the particle
radii are small enough compared to the incident wavelength to treat the 
particles as electric dipoles with a moment given by $\mathbf{p} = 
\epsilon\epsilon_{0}\alpha\mathbf{E}$, with complex electric polarisability 
$\alpha= \alpha' + i \alpha''$ and total field $\mathbf{E}$. We used $k = n 
k_0 = n 2\pi / \lambda$ for the wave-number of the incident laser light, close
to the gold plasmon resonance in water, which gives a wavelength of $\lambda
\approx 395\ \mathrm{nm}$. For these values $\alpha' \approx 1\times 10^{-21}\ 
\mathrm{m}^3$ and $\alpha'' \approx 2\alpha'$. For an incident electric field 
$\mathbf{E}_0$, the field intensity considered here, $(1/2)\epsilon_{0}cn 
|\mathbf{E}_0|^2 \sim 10^{9}\ \mathrm{W/m}^2$, could be achieved by focusing a 
$0.1\ \mathrm{W}$ laser onto a $(10\ \mu\mathrm{m})^{2}$ region. We have used 
$c$ to represent the speed of light and $\mathbf{u}_z$ for the unit vector in 
the direction perpendicular to the $xy$ plane. According to Eq. 
(\ref{force_field}), the energy from the light shone on a gold nanoparticle is 
typically $U \equiv 2I\frac{n}{c}\alpha'$. For the $50\ \mathrm{nm}$-radius gold 
particles, experimentally feasible laser intensities lie in the $U \in [0-10^2]\ 
k_BT$ range at $T = 300\ \mathrm{K}$, with $U \simeq 1.7 k_BT$ for the 
aforementioned $0.1\ \mathrm{W}$ laser.

Our flexible dumbbells consist of a chain of $N$ spheres connected with standard
finitely extensible nonlinear elastic (FENE) bonds \cite{Warner1972}. In addition,
we have included semiflexible potentials to simulate the rigidity of the polymer 
\cite{Kierfeld2004} and Weeks-Chandler-Andersen (WCA) interactions among the 
beads to account for excluded volume effects \cite{Andersen1971}.
We chose a persistence length of $l_p = 50\ \mathrm{nm}$, typical of
double-stranded DNA, though changes in the range of $50$--$5000\ \mathrm{nm}$
did not significantly alter the dispersion of the $100$--$1000\ 
\mathrm{nm}$-length chains of interest to this study (mean square displacements
changed by less than a factor of two). The beads at  the two ends of the chain
represent gold nanoparticles subject to the optical  forces in Eq.
(\ref{force_field}) (see Fig. \ref{fieldfig}). Most of the simulations added the
effect of light scattering off the end particles \cite{Delgado2018}. In these
cases we included an extra coat of transparent material (represented with dotted
lines in the figure) mimicking the experiments in Ref. \cite{Honda2011}, which
prevented the gold spheres from coming too close to each other as that would
make the approximations involved in the calculation of the scattering unsound.
If the coat were removed, the equations of both the forces and the optical
scattering would have to be modified for small separations, and the possibility
of particle merging would have to be taken into account.

For the time scales of interest here, Brownian dynamics (overdamped Langevin 
equations of motion) accurately describe the fluctuating motion of our nanoscale 
dumbbells \cite{Fixman1978}. Using $\mathbf{R}$ to represent a vector of all the 
coordinates, a realisation of the particle trajectories corresponds to the 
solution of the following stochastic differential equation \cite{Doyle2005}, 
which we integrated numerically with an Euler-Maruyama scheme.
\begin{equation}
  \label{Brownian_dynamics}
  d\mathbf{R} = \mathbf{M F}\ dt + \sqrt{2k_BT} \mathbf{B}\ d\mathbf{W}.
\end{equation}
The first term on the right stands for the deterministic flow, with the mobility
tensor $\mathbf{M}$ multiplying the net force $\mathbf{F}$, which includes the
optical forces in Eq. (\ref{force_field}), the WCA interaction among beads, the
FENE bonds and the semiflexible potential. The last term incorporates the effect of
thermal fluctuations through a random force. $\mathbf{B}$ satisfies the relation
$\mathbf{B B}^T = \mathbf{M}$, $d\mathbf{W}$ is the Wiener process, and $k_BT$
is Boltzmann's constant multiplied by the temperature.

By using the Rotne-Prager-Yamakawa tensor \cite{Rotne1969,Yamakawa1970} as a 
mobility matrix, you can model hydrodynamic interaction among the particles. The 
$\mathbf{B}$ matrix was calculated with the usual algorithm for Cholesky
decomposition \cite{Hammerlin2012}. In some of our simulations, though, we
disregarded  hydrodynamic interactions altogether. In that case, $\mathbf{M}$
equals the identity  matrix multiplied by $\gamma^{-1}$, the inverse of the
friction coefficient, and  the calculation of $\mathbf{B}$ becomes trivial.
Comparing the latter purely Brownian simulations to the other results allowed us
to isolate the effects due to hydrodynamics.

In addition to hydrodynamic interactions, a realistic simulation of the moving 
dumbbells must include the secondary forces arising from the light scattered by 
the gold particles. Besides the incident field $\mathbf{E}_0$, the total
electric field $\mathbf{E}$ at the position of particle $i$, $\mathbf{r}_i$ must 
then include a contribution from scattering proportional to the Green's function 
propagator $\mathbf{G}(r_i,\ r_j)$ at the position of other particles
\cite{Novotny2006} 
multiplied by the total field at those locations $r_j$, that is to say,
\begin{equation}
\label{totalfield}
\mathbf{E}(\mathbf{r}_i) = \mathbf{E}_0(\mathbf{r}_i)
                           + k^2 \sum_{j \neq i}
                             \mathbf{G}(\mathbf{r}_i, \mathbf{r}_j)\ 
                             \alpha \mathbf{E}(\mathbf{r}_j).
\end{equation}
By solving the equation above numerically for $\mathbf{E}(\mathbf{r}_i)$
\cite{Delgado2018} and inserting the result into the expression for the force,
we obtain the force on particle $i$ \cite{Luis-Hita2016}.
\begin{equation}
  \label{optical_force}
  \mathbf{F}_{opt}(\mathbf{r}_i)
    = \frac{\epsilon\epsilon_0}{2} \Re(\alpha \mathbf{E}(\mathbf{r}_i)
                                       \cdot \nabla \mathbf{E}^*(\mathbf{r}_i)).
\end{equation}
This means that component $\mu$ of the average force on particle $i$ equals
\begin{equation}
  (F_{opt}(\mathbf{r}_i))_{\mu}
    = \frac{\epsilon\epsilon_0}{2}
      \Re\left(\alpha \sum_\nu E_\nu(\mathbf{r}_i)
               \frac{\partial}{\partial x_\mu} 
               E_\nu^*(\mathbf{r}_i)\right).
\end{equation}
Therefore, Eq. (\ref{optical_force}) replaces Eq. (\ref{force_field}) when the
simulations take light scattering into account. As we will show below,
scattering alters the motion of the dumbbells qualitatively.

\section{Results}

The schematic representation of the optical field (\ref{force_field}) shown in 
Fig. \ref{fieldfig} indicates the direction of the forces acting on a single 
nanoparticle due to the incident laser beams. The particles (and, one might 
expect, chains much shorter than the wavelength) follow the lines of force and 
experience enhanced dispersion \cite{Albaladejo2009} moving from one unstable 
node to another. Attaching two gold spheres alters the picture, because they may 
become entrapped by vortices for certain values of the parameters, as 
illustrated in Fig. \ref{fieldfig}. For example, \textit{when we neglect 
hydrodynamic interactions}, relatively rigid 10-bead chains display enhanced 
dispersion when the laser wavelength equals 100 times the radius of the gold 
particles, but remain trapped, rotating in a vortex, when the wavelength 
approximately doubles the length of the dumbbell (as in Fig. \ref{fieldfig}b), 
even when the program includes the effects of light scattering ($r = 50\ 
\mathrm{nm}$, $U = 10\ k_BT$ and persistence length $l_p = 1000\ r$). Similarly, 
dispersion plummets when the wavelength approaches the length of the chain, 
which remains stuck between two vortices, with both ends rotating in the same 
direction (Fig. \ref{fieldfig}c). Not much changes when we reduce the 
persistence length to that of dsDNA, except that at the larger values of the 
wavelength ($\lambda=50$--$100$) the chains show a greater tendency to fold in 
half and remain circling a vortex, but the secondary scattering eliminates this 
effect.

\begin{figure}
  \includegraphics[width = 0.8 \columnwidth]{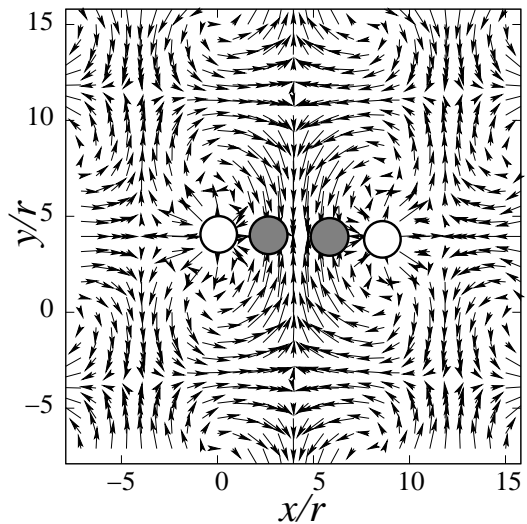}
  \caption{\label{dumbbell_scattering}Dumbbell sitting in a saddle due to the
  combined effect of the incident laser field and light scattering off the gold
  particles at the ends ($\lambda = 7.9\ r$, $U = 10\ k_BT$). Note that the
  optical forces stretch the chain.}
\end{figure}

Notwithstanding the interest of this dynamical behaviour, here we have chosen
to concentrate on realistic values of the parameters, considering current
experimental technology. As mentioned above, the optical field for $r = 50\ 
\mathrm{nm}$ gold particles has a wavelength $\lambda \approx 4 r$. This
implies that dumbbells short enough to lie in a vortex feel strong forces due
to light scattering off the ends, which significantly modify the force
landscape. For instance, 4-bead dumbbells tend to move to a nearby saddle point
and sit there (Fig. \ref{dumbbell_scattering}). We observed the same behaviour
when the intermediate FENE chain was approximated by means of a single harmonic
bond.

\begin{figure}
  \includegraphics{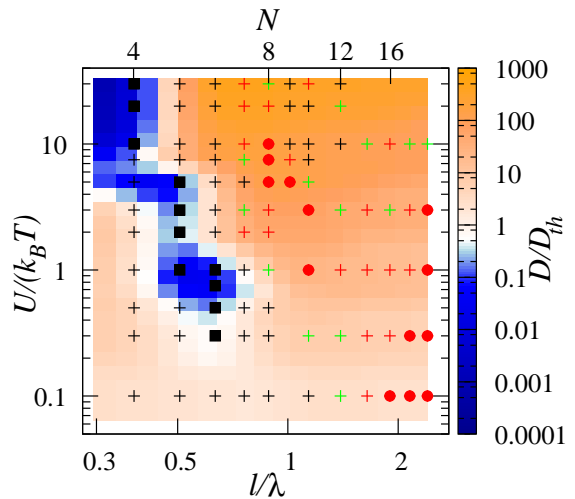}
  \caption{\label{phase_diagram}Dispersion coefficient $D$ divided by thermal 
  diffusion $D_{th}$ for different values of the laser intensity and dumbbell 
  length. The colour scale values were obtained by interpolation. Points mark 
  the values of the parameters for which simulations were performed. See Fig. 
  \ref{phase_diagram_explanation}) for an explanation of the meaning of the 
  differences among points. $D_{th}$ was calculated numerically for each length 
  by switching off the laser forces.}
\end{figure}

\begin{figure}
  \includegraphics[width=\columnwidth]{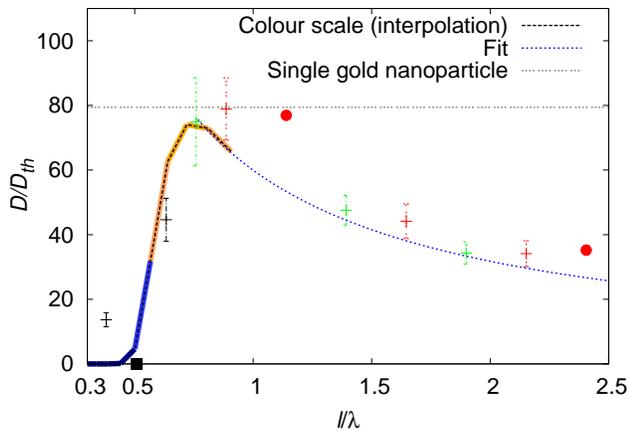}
  \caption{\label{phase_diagram_explanation}Dispersion coefficient $D$ divided 
  by thermal diffusion $D_{th}$ for different dumbbell lengths. This figure 
  presents a horizontal section of Fig. \ref{phase_diagram} for $U = 3\ k_BT$. 
  Red points lie above a fitted function $D_{\mathrm{fit}} \propto U^{0.84}/l$ 
  and black values below. Green points coincide with the fit within their margin 
  of error. Squares correspond to trapped dumbbells ($D/D_{th} < 0.1$) and 
  circles to dumbbells with dispersion coefficients greater than 
  $D_{\mathrm{fit}}$ by at least 20\%. We also show the value of $D/D_{th}$ for 
  a single gold nanoparticle in the same optical force field. The fit resulted 
  from approximating all the simulations shown in Fig. 3, not just the ones
  shown here.}
\end{figure}

\textit{Adding hydrodynamics} into the mix induces coupling among the bead 
displacements, and this has a significant effect on the dynamics, as the mutual 
drag pulls the dumbbells out of the vortices. The hydrodynamic interaction 
between two nearby nanoparticles in a vortex lattice has been analysed in Ref. 
\cite{Delgado2018}. Fig. \ref{phase_diagram} charts the scaled dispersion 
coefficient $D$ of a dumbbell as a function of two parameters: the laser 
intensity and the chain length, with $D = \langle \Delta x^2 + \Delta y^2 
\rangle / (4 \Delta t)$, and $\Delta x = x(t + \Delta t) - x(t)$ and $\Delta y = 
y(t + \Delta t) - y(t)$, and $D_{th}$ representing the thermal diffusion 
coefficient with the optical fields switched off. The colour scale values shown 
in Fig. \ref{phase_diagram} were calculated by interpolation. We can still find 
a region, indicated by the blue area in Fig. \ref{phase_diagram}, in which the 
dumbbells remain trapped on saddle points according to the mechanism shown in 
Fig. \ref{dumbbell_scattering}. Close to this region, approximately for $\lambda 
\approx l$, we observe a strong enhancement of dispersion (marked with red 
circles in Fig. \ref{phase_diagram}) as also evidenced in Fig. 
\ref{phase_diagram_explanation}. For $\lambda > l$, $D$ scales very roughly as 
$D/D_{th} \propto U^{0.84}/l$ (corresponding to points marked with crosses, see 
also Fig. \ref{phase_diagram_explanation}).

\begin{figure}
  \includegraphics[width = \columnwidth]{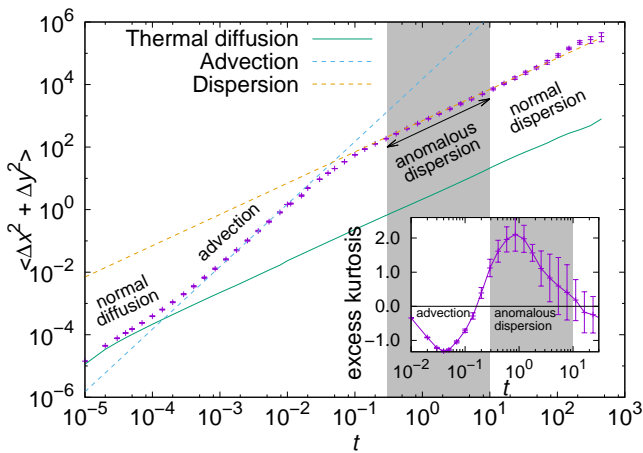}
  \caption{\label{diffusion_stages}Mean square displacement in units of 
  $\sigma^2$ for an $N = 8$ dumbbell at $U = 7.5\ k_BT$ (points) showing the 
  transition from thermal diffusion at short times to enhanced dispersion at 
  long times (measured in Brownian time units $\sigma^2/D_{th}$). Anomalous 
  Brownian motion was observed for times in the range $0.3$--$10$ (shaded area). 
  The inset shows that the excess kurtosis for the displacement distribution 
  becomes positive in the same range.}
\end{figure}

Fig. \ref{diffusion_stages} displays the mean square displacement of a moving 
dumbbell on the $xy$ plane for different time scales. The general shape of the 
curve coincides with that of a single gold particle, with Brownian motion at 
short and long time intervals connected by means of an advective regime at 
intermediate time scales. Below the mean free time of molecular collisions (not 
shown in the figure) the physical system would move in a ballistic regime, with 
the mean square displacement proportional to the time squared. At larger time 
scales, we find thermal diffusion. Because the optical forces push the gold 
nanoparticles from one saddle node to the next, the dumbbell then encounters an 
advective regime, where the mean square displacement scales once again with 
$t^2$. Finally, at long times the motion of the dumbbell behaves like diffusion, 
but with a larger diffusion coefficient. In some cases, including that of Fig. 
\ref{diffusion_stages} ($N = 8$, $U = 7.5\ k_BT$), we found anomalous Brownian 
motion for a range of times, in contrast to the case of single nanoparticles. 
This non-Gaussian character of the displacements cannot be seen in the graph of 
the mean square displacement, but is confirmed by the inset in Fig. 
\ref{diffusion_stages}, which plots the excess kurtosis for the displacement 
distribution and reveals that it becomes positive in the range of anomalous 
Brownian motion. If we disregard hydrodynamic interactions, the anomalous 
character of the motion disappears.

For strict thermal diffusion, the dumbbell $D_{th}$ decreases with the length of 
the chain, but for the phenomenon studied here, the dispersion in some cases 
approached that of single nanoparticles (Fig. \ref{phase_diagram_explanation}).

\begin{figure}
  \hspace*{-0.5cm}
  \includegraphics[width=1.05\columnwidth]{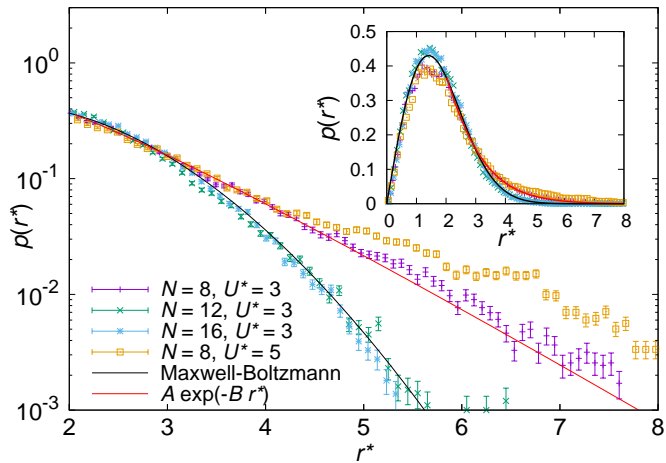}
  \caption{\label{step_pdf}Probability density function tails for the 
  displacement of diffusing dumbbells on a semi-logarithmic scale. The 
  horizontal axis represents a scaled step
  $r^* = r_{\Delta t}/\sqrt{(U^*/n)\Delta t}$, with $\Delta t = 100\ \sigma^2/D_{th}$.
  The inset shows the distributions on a linear scale. The number of beads $N$ 
  includes the gold nanoparticles. $U^*$ means $U/(k_BT)$. This function 
  represents the dumbbell contribution to the radial van Hove function for a 
  fixed time delay $\Delta t$.}
\end{figure}

The probability distribution function for the size of a step in a
two-dimensional random walk follows a Maxwell-Boltzmann curve
\begin{equation}
P(r_{\Delta t})
  = \frac{r_{\Delta t}}{\sigma^2}e^{-r_{\Delta t}^2/(2 \sigma^2)},
\end{equation}
with $r_{\Delta t}$ a shorthand for $\|\mathbf{r}(t + \Delta t) - 
\mathbf{r}(t)\|$. This function may be viewed as the self contribution to the 
radial van Hove function for a time delay equal to $\Delta t$. Single 
nanoparticles on the vortex lattice follow a jagged version of this distribution 
and some values of the parameters for dumbbells produce similar distributions 
($N = 12,\ 16$, $U^* = 3$ in Fig. \ref{step_pdf}). Contrast the trend along the 
tail of the distributions to the behaviour near regions marked with red dots in 
Fig. \ref{phase_diagram} ($N = 8$, $U^* = 3,\ 5$), which fall off exponentially, 
indicating the anomalous (though still Brownian) nature of the diffusive motion 
in these cases \cite{Wang2009}. The data in Fig. \ref{step_pdf} were all 
obtained from time lags $\Delta t$ for which the in-plane dispersion coefficient 
$D_{th}$ had reached a plateau. For larger values of the time lag, the 
distribution reverted to the standard Gaussian Maxwell-Boltzmann statistics 
\cite{Wang2012,Chubynsky2014}. Accordingly, despite the superficial resemblance 
to L\'evy walks \cite{Zaburdev2015}, the long-time dynamics follow normal random 
walk statistics. A comparison with simulations without scattering revealed that 
scattering, in fact, postpones the arrival of Gaussian distributions.

\begin{figure}
  \includegraphics[width=\columnwidth]{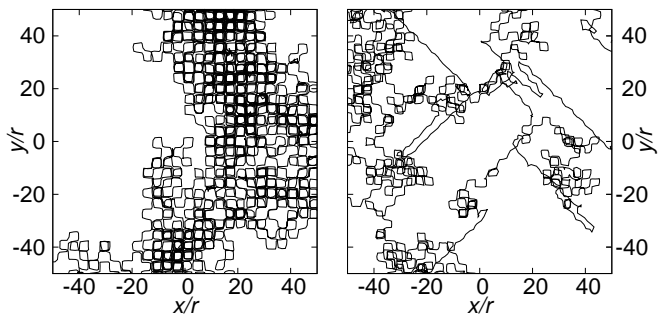}
  \caption{\label{compare_trajectories}Centre of mass trajectories of a dumbell
  in an optical vortex lattice ($U = 7.5\ k_BT$). The $N = 5$ dumbbell on the
  left displays normal (Gaussian) enhanced Brownian displacements at long times,
  while the $N = 8$ dumbbell on the right has a non-Gaussian displacement
  distribution function due to the higher probability of a long displacement
  along one of the $x = \pm y$ diagonals caused by hydrodynamic drag.}
\end{figure}

The higher-than-normal probability of a large displacement, compared to Gaussian
statistics, emerges from a tendency to move along the $x = \pm y$ diagonal
directions in the lattice due to hydrodynamic interactions \cite{Delgado2018}
and can be spotted easily by comparing the trajectory of the Gaussian
displacements on the left of Fig. \ref{compare_trajectories} to that of an
anomalous Brownian process (Fig. \ref{compare_trajectories}, \textit{right}).
In contrast, single nanoparticles do not prefer to move along diagonals. They 
simply move to the next saddle node and randomly choose one of two opposite 
directions, alternating up/down with left/right.

\section{Limitations of the present approach}

The fact that optical absorption induces heating in the gold nanoparticles
\cite{Seol2006} places an upper bound on the intensity of the laser source. The
phenomena described in our paper would obviously change dramatically if the
heating caused the water to boil around the particles. Previous research has
reported such a formation of vapour bubbles \cite{Govorov2007,Yeschenko2016}.
The water in contact with the metal sphere would increase its temperature by an
amount proportional to the square of the sphere radius $r$ and the laser
intensity $I$ \cite{Govorov2007},
\begin{equation}
  \Delta T_{max} \propto r^2 I.
\end{equation}
The data provided in \cite{Govorov2007} allow us to estimate a temperature 
increase $\Delta T \approx 120\ \mathrm{K}$ for our $10^9\ \mathrm{W/m}^2$ laser 
and $50\ \mathrm{nm}$-radius sphere. However, this analytical solution neglects 
several effects due to the temperature dependence of optical and thermal 
properties of the gold particles, which all lead to a overestimation of $\Delta 
T$ in the case of gold nanoparticles in water \cite{Sivan2017}. However, the 
combined heating effect of several particles would indeed lead to much higher 
temperatures \cite{Govorov2007,Yeschenko2016}, and this sets a limit on the 
maximum volume fraction for which the present analysis holds. Nevertheless, 
temperature gradients around the nanoparticles might still have an effect on 
their motion due to thermophoresis \cite{Piazza2004}, even below the boiling 
temperature.

Lastly, we would like to emphasise that we have focussed on the dynamics of a
single dumbbell. At high enough concentrations of suspended dumbbells,
hydrodynamic interactions would distort their motion, and the flows created by a
moving dumbbell would affect the others, giving rise to collective dynamics, as
proven in a previous work \cite{Delgado2018}.

\section{Conclusions}

Nanoscale dumbbells, created by attaching two 50-$\mathrm{nm}$-radius gold
spheres by means of a polymer chain, enrich the dynamics predicted by
Albadalejo \textit{et al.} for nanoparticles in an optical vortex lattice
\cite{Albaladejo2009}. Not only might they experience enhanced dispersion for
large values of the laser intensity, but they can be trapped in a fixed position
or be made to rotate (when $l/\lambda \approx 0.5$). The different behaviours
depend on the ratio of the chain length to the wavelength $l/\lambda$ and the
intensity of the incident lasers creating the vortex field.

As displayed in Fig. \ref{phase_diagram}, relatively modest changes in the size
of the chains or the laser intensity may lead to huge variations of the
dispersion of dumbbells along the plane spanned by the direction of the lasers,
and this provides a way to guide, filter or trap the dumbbells by tuning the
intensity of the lasers.

\section{Acknowledgements}

We would like to thank J. Luis-Hita for his code to calculate the optical
forces. This research was funded by the Spanish MINECO project
FIS2017-86007-C3-1-p, the MINECO and European Regional Development Fund project
FIS2015-69295-C3-3-P, Explora Ciencia FIS2013-50510-EXP, and the Basque
Department of Education project PI-2016-1-0041.


\end{document}